\documentclass[12pt,a4paper]{article}
\usepackage{amsmath,amssymb}
\usepackage{epsfig}
\usepackage{citesort}
\usepackage{times}
\usepackage{calc}
\newcommand{\bra}{\langle}
\newcommand{\ket}{\rangle}
\newcommand{\ul}[1]{#1} 
\date{}
\makeatletter
\renewcommand{\section}{\@startsection
{section}
{1}
{0mm}
{2\baselineskip minus \parskip}
{1\baselineskip minus \parskip}%
{\normalfont\normalsize\bfseries}}%

\makeatother
\begin{document}
\begin{center} 

\vspace*{1cm}

{\huge Charge Transport in Polymer Ion Conductors: a Monte Carlo
  Study\\}
\vspace{1cm}
{O. D\"urr$^{1}$, W. Dieterich$^{1}$ and A. Nitzan$^{2}$\\
${}^{1}${\it Fachbereich f\"ur Physik, Universit\"at Konstanz, D-78457
  Konstanz, Germany\\
${}^{2}$School of Chemistry, The Sackler Faculty of Science, Tel Aviv
University, Tel Aviv 69978, Israel}}
\end{center}

\begin{abstract}
Diffusion of ions through a fluctuating polymeric host is studied both
by \ul{Monte Carlo simulation} of the complete system dynamics and by
\ul{dynamic bond percolation} (DBP) theory. Comparison of both methods
suggests a multiscale-like approach for calculating the diffusion
coefficients of the ions.
 
\end{abstract}
\section{Introduction}
Chain polymers carrying electro-negative atoms (e.g. oxygen or
nitrogen) in their repeat unit can act as solvents for certain salts.
Well-known examples are Li-salts dissolved in polyethylene-oxide (PEO).
At temperatures sufficiently above the glass-transition temperature
these polymer-salt solutions show significant DC ionic $(Li^{+})$
conductivities. Such "polymer electrolytes" offer widespread
applications in batteries, sensors and fuel cells. From the scientific
point of view, an important goal is to improve our
understanding of the 
electrical conduction mechanism in polymer electrolytes, and in
particular to elucidate  the interplay between ion \ul{diffusion} and the
polymer network dynamics.


Dynamic Monte Carlo (MC) simulation  of the diffusion coefficient of few
particles (ions) 
in a rearranging environment of polymer chains 
is hampered by the need to move every monomer (polymer bead or ion)
with the same probability. Therefore most of the computational time is spent
moving the polymeric host without affecting the ionic configurations. 
On the other hand,
earlier studies indicate that important features of the ion diffusion
within a dynamical matrix of chain molecules can be described
by a more coarse-grained model\cite{PhysicaA}. 
The idea is to map the diffusion process onto DBP-theory \cite{DRU85,Dru88}  
and to determine the central quantity entering this theory, the
renewal time distribution $\psi(t)$, 
from the time dependence of the local occupational correlation
function due to the polymer chain dynamics.
In this communication we perform tests of such a procedure for a
hard-core lattice gas and for tracer diffusion in systems of athermal
\ul{lattice polymers}.
It turns out that the DBP-concept compares favourably with simulations
of the complete system dynamics and thereby saves about
one order of magnitude in computer time. 

\section{Hard-core \ul{lattice gas}}
In order to explain our procedure we treat the simple case of a
hard-core lattice gas, i.e.
point particles moving on a lattice with no interaction besides hard
core repulsion. 
The tracer diffusion constant $D(c)$ as a function of concentration
$c$ is known to a high degree of accuracy via dynamic pair approximations
\cite{Na80,Ta83,Di1} which yield a tracer correlation factor 
\begin{equation}
  \label{eq:1}
  f(c) :=\frac{D(c)}{D_{0} (1-c)} =
  \frac{1+\langle\cos\Theta\rangle}{1-[(3-2c)/(2-c)]\bra \cos\Theta
  \ket },
\end{equation}
with $\bra \cos\Theta \ket  \approx -0.209$ for a simple cubic lattice.

Now we consider a lattice with static disorder in which randomly
chosen sites are blocked and thus not accessible for the tracer particle.
For this standard percolation problem we denote the mean square
displacement of the tracer by $\bra r^{2}(t) \ket_{0}$.
If on the other hand the blocked sites are globally and
instantaneously rearranged according to an arbitrary 
waiting-time distribution $\psi(t)$, generalized DBP-theory yields the
following diffusivity at zero frequency in $d=3$ dimensions \cite{Dru88}
\begin{equation}\label{eq:2}
D=\frac{1}{6}\frac{\int_{0}^{\infty}dt \psi(t)\langle
  r^{2}(t)\rangle_{0}}{\int_{0}^{\infty}dt\,t\psi(t)}
\end{equation}

Our aim is to map the complete system of coupled ions and lattice
chains  onto this
coarse-grained DBP-model. While $\bra r^{2}(t) \ket_{0}$ can be
obtained in a straightforward manner from simulations of ion diffusion
in the frozen network, the determination of $\psi(t)$ requires more
explanation. 
Within the spirit of previous work on the hard-core lattice gas 
we propose to determine $\psi(t)$ from the local occupational correlation
function $\bra n_{i}(t) n_{i}(0)\ket$ of a site i adjacent to a fixed
tracer position. Let us introduce the probability $\Phi(t)$ that no
renewal takes place within the time interval $[0,t]$ after a previous
renewal at an 
arbitrary $t_{0}<0$. 
The joint probability $\bra n_{i}(t) n_{i}(0) \ket$ 
that the lattice point $i$ is occupied at $t=0$ and at $t$ (not
necessarily by the same particle) consists of two distinct contributions. The
first one is the probability that $i$ is occupied at $t=0$ and that no
renewal occurs until $t$, which is given by $c \; \Phi(t)$. The second
contribution describes the situation where one or more renewals have taken
place until $t$. The corresponding probability is given by
$c^{2}(1-\Phi(t))$. Hence $\Phi(t)$ is related to  
$\bra n_{i}(t) n_{i}(0) \ket$ via
\begin{equation}
  \label{eq:3}
  \Phi(t)=\frac{\langle n_{i}(t)n_{i}(0)\rangle -c^{2}}{c(1-c)}
\end{equation}
According to \cite{Dru88}, $\psi(t)=\Phi''(t) {\bar{\lambda}}^{-1}$
where $\bar{\lambda}=\int_{0}^{\infty} t \psi(t) dt$ denotes the mean
renewal time.  Equation (\ref{eq:2}) thus can be rewritten as 
\begin{equation}
  \label{eq:4}
  D=\frac{1}{6}\frac{\int_{0}^{\infty}dt\Phi''(t)\langle
  r^{2}(t)\rangle_{0}}{\int_{0}^{\infty}dt\,t\Phi''(t)}
\end{equation}
where we obtain $\Phi(t)$ from (\ref{eq:3}) by means
of MC simulations for $\bra n_{i}(t)n_{i}(0) \ket$. 
The resulting $\Phi(t)$ can be fitted with 
sufficient accuracy  by a combination  of $P$ exponential functions,
\begin{equation}
  \label{eq:5}
  \Phi(t) = \sum_{i=1}^{P} a_{i} \exp(- \lambda_{i} t)
\end{equation}
On the other hand, $\langle r^{2}(t)\rangle_{0}$ can by expressed in terms of the
simulated mean square displacement $\bra r_{n}^{2} \ket_{0}$ of a tracer
particle making $n$ steps in a frozen network  via   
\begin{equation} 
  \label{eq:6}
  \langle r^{2}(t)\rangle_{0} = \lim_{N \rightarrow \infty}
  \sum_{n=0}^{N} \frac{(\omega_{0} t)^{n}}{n!}  \exp(-\omega_{0}t)
  \langle r^{2}_{n} \rangle_{0} 
\end{equation}
where $\omega_{0}$ is the attempt frequency of the tracer.
Inserting (\ref{eq:5}) and (\ref{eq:6}) into (\ref{eq:4}) and
carrying out the integration one gets
\begin{equation}
  \label{eq:7}
  \frac{D}{\omega_{0}} = \frac {1}{6} \lim_{N \rightarrow \infty}\sum_{i=1}^{P}
  (\frac{\lambda_{i}}{\omega_{0}})^{2} a_{i} \sum_{n=0}^{N} \frac{\bra r^2_n
  \ket_{0}}{(\lambda_{i}/\omega_{0}+1)^{n+1}}
\end{equation}
Independent calculations for $c
\rightarrow 1$ showed that in order to keep the error caused by the
finiteness of $N$ and $p$ below the statistical error, values of $P=3$ and
$N=40$ were sufficient.  
Note that the main decay of $\Phi(t)$ is roughly represented
by a single exponential with decay rate
$\lambda_{0}=-(d\Phi/dt)_{t=0}=(5/6)\omega_{0}$.

\begin{figure}[tb]
  \begin{center}
     \epsfig{file=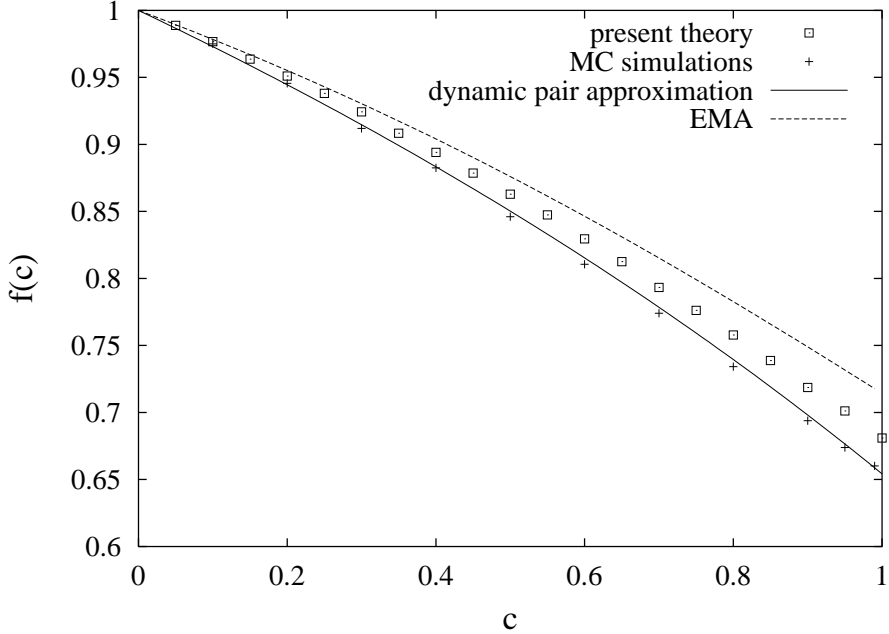,width=0.8\linewidth,angle=0} 
    \caption{Test of DBP-theory against simulations of the complete
      system and the dynamic pair approximation equation (\ref{eq:1}).
      The quantity shown is the tracer correlation function $f(c)$ as
      a function of concentration. Also plotted are results obtained
      recently form a many-particle effective medium approximation\cite{Gra90}.} 
    \label{fig:1}
  \end{center}
\end{figure}

In Figure (\ref{fig:1}) we compare the results of this theory with MC
simulations (see also \cite{BRU73}) and the dynamic pair approximation
equation (\ref{eq:1})
and also with an existing effective 
medium approximation (EMA) for a percolating lattice renewing itself
with only one rate constant $\lambda=5/6 \omega_{0}$ \cite{Gra90}.
As one can see, using more than one rate constant in the representation
of $\Phi(t)$ according to (\ref{eq:5})
improves the results significantly.
This observation suggests to apply the generalized DBP-theory to the
situation described in the next section where the non-exponential
character of the \ul{renewal process} is even more pronounced.

\section{Results for athermal chains and conclusion}

In the case of a hard-core lattice gas the procedure described  bears
no computational advantage over existing methods but was merely
considered as a test case.
The situation
changes, however, when we apply our approach to ion diffusion in a
polymer network. No analytic
approximations for $D(c)$ of a quality similar to (\ref{eq:1}) are
available then, and full simulations of $D$ become much more demanding
because of the internal degrees of freedom of the host molecules and
the larger statistical errors connected with the small concentration
of tracer particles. 

Let us consider, for example, a system where the chains are made of
ten beads, assigned to lattice sites, and linearly connected via
nearest-neighbour bonds. Apart from site exclusion, which mimics a
hard--core repulsion, 
no explicit interactions between beads are assumed.
The size of the simulation box is $L=10$ and periodic boundary
conditions in all directions are employed. For the chain dynamics we
use the standard algorithm as described in Ref. \cite{Ve62,Hi75,Kr88}, while
point-particles individually perform nearest-neighbour hops. 

\begin{figure}[tb]
  \begin{center}
     \epsfig{file=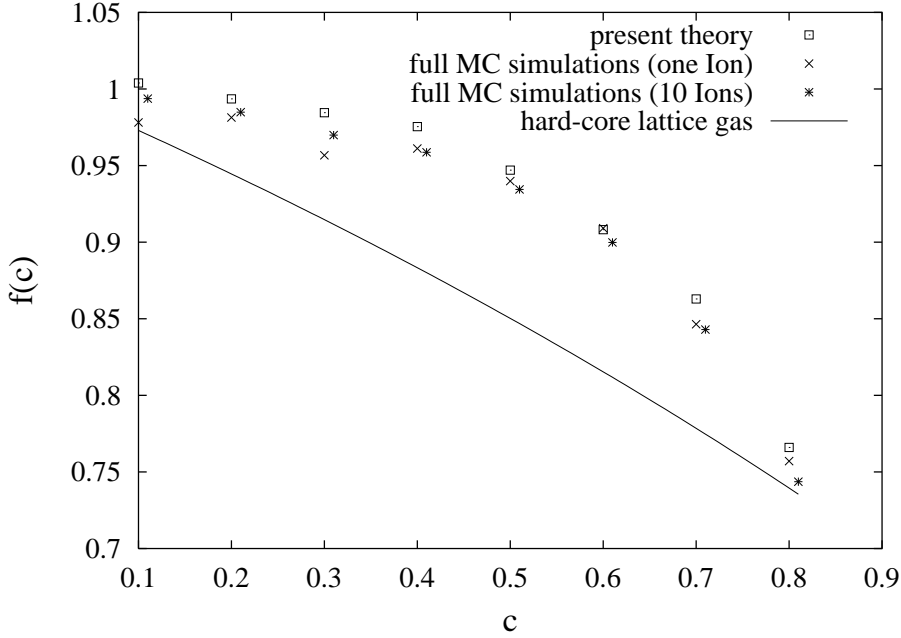,width=0.8\linewidth,angle=0} 
     \caption{Comparison of the present theory with full MC simulation
       using ion concentrations $c_{\text{ion}}=10^{-3}$ (one ion)
         and $c_{\text{ion}}=10^{-2}$ (10 ions).
         For comparison the result of equation (\ref{eq:1}) for a
         hard-core lattice gas is also included.}
    \label{fig:2}
  \end{center}
\end{figure}

To apply our theory we again determine $\Phi(t)$ in analogy to the
hard-core lattice gas, see (\ref{eq:5}), which is now highly
non-exponential, indicating the importance of temporal correlations in
the associated renewal processes. A superposition of three
exponential functions in (\ref{eq:5}), however, still gives a good description
of $\Phi(t)$.  
Mean-square displacements 
$\bra r_{n}^{2} \ket_{0}$ of ions within a frozen network  of chains 
are determined using $N=200$ steps,
and the diffusion constant is calculated as before via equation (\ref{eq:7}).
In Fig.(\ref{fig:2}) we compare our results for the tracer correlation
factor as a function of the total concentration of occupied sites, c,
with full MC simulations. Due to the connectivity of chains blocking
effects at a given $c$ on average are reduced and the \ul{correlation
factor} $f(c)$ is larger than in the case of point particles, at least
as long as $c \lesssim 0.8$.
More important from the computational point of view is the fact that
DBP-theory very well agrees with the full simulations, but saves about
one order of magnitude in computer time. Mapping that problem onto a
coarse-grained  DBP-model thus appears to constitute an accurate and
efficient method for investigating diffusion through a dynamic host
of chain molecules. To support this conclusion further work is being
carried out, in particular under varying chain lengths.

\end{document}